\documentclass[twocolumn,
 amsmath,amssymb,
 aps,
prb,
]{revtex4}

\usepackage[squaren,Gray]{SIunits}

\usepackage{graphicx}
\usepackage{dcolumn}
\usepackage{bm}

\begin{document}

\title{
Contrasting influence of charged impurities on transport and gain in terahertz quantum cascade lasers
}

\author{Thomas Grange}
\email{thomas.grange@neel.cnrs.fr}
\affiliation{Walter Schottky Institut, Physik Department, Technische Universit\"{a}t M\"{u}nchen, D-85748, Garching, Germany}
\affiliation{Universit\'e Grenoble-Alpes, F-38000 Grenoble, France}
\affiliation{CNRS, Institut N\'{e}el, "Nanophysique et semiconducteurs" group, F-38000 Grenoble, France}

\date{\today}

\begin{abstract}
Transport and gain properties of a resonant-phonon terahertz quantum cascade laser are calculated using nonequilibrium Green's functions. 
Impurity scattering is shown to be responsible for contrasting nonlinear effects in the transport and the gain properties.
For typical doping concentrations, the current density is found to be weakly sensitive to the impurity scattering strength. In contrast, the calculated gain is found to be very sensitive to the impurity scattering strength.
This difference is attributed to the strong momentum dependence of the long-range coupling to charged impurities. Small-momentum impurity scattering is shown to be responsible for an incoherent regime of resonant tunneling processes.
These new insights into the crucial role of impurity scattering open a new route of improvement of terahertz quantum cascade lasers by engineering of the doping profile.


\end{abstract}

\maketitle


Since the discovery of terahertz (THz)  quantum cascade lasers (QCLs), great efforts in optimizing these devices have been made with the ultimate goal of developing high-power THz lasers operating at room temperature \cite{Kohler02,williams07,vitiello2015quantum,li2014terahertz}.
Among the THz QCL design parameters, the influence of the doping concentration on the performance of these devices has been studied experimentally in various THz QCLs \cite{liu2005effect,ajili2006doping,benz2007influence,andrews2008doping,deutsch2013dopant}. Theoretically the elastic scattering of electrons by ionized charged impurities has be shown to play a key role \cite{callebaut2004importance,banit2005self,nelander2008temperature,nelander2009temperature}.
However, a conclusive understanding of the influence of charged impurities on the transport and gain properties is still lacking.
Experimentally, the paradox is the following: (i) the current density is observed to evolve almost linearly with the doping density, suggesting a minor role of charged impurity scattering (CIS); (ii) the output power does not evolve linearly, but rather saturates, suggesting an important influence of CIS. An optimum for the terahertz amplification is observed at a doping concentration at which the transport properties still evolve linearly.
This paradoxal behavior has not been explained by any of the various theoretical studies reported so far \cite{pereira2004controlling,jovanovic2005mechanisms,leuliet2006electron,jirauschek2007comparative,jirauschek2009monte,Kumar09b,terazzi2010density,dupont2010simplified,wacker2011unraveling,carosella2012free,matyas2013role,ndebeka2013dopant,jirauschek2014modeling,agnew2015efficient,franckie2015impact,
lee2002nonequilibrium,wacker2002gain,lee2006quantum,kubis09,schmielau2009nonequilibrium,yasuda2009nonequilibrium,kubis2010design,wacker2013nonequilibrium}.

Here we calculate the transport and gain properties of a resonant-phonon THz QCL using the formalism of nonequilibrium Green's functions (NEGF) \cite{lee2002nonequilibrium,wacker2002gain,lee2006quantum,kubis09,schmielau2009nonequilibrium,yasuda2009nonequilibrium,kubis2010design,wacker2013nonequilibrium}. Simulations are done for varying CIS strengths and varying doping concentrations.
We demonstrate that CIS has a markedly different effect on the transport and the gain properties. 
For typical experimental doping densities, CIS is shown to be a major source of saturation of the gain with increasing doping density. In contrast, the current densities evolve almost linearly for the same doping densities.
Yet CIS is shown to be responsible for the incoherent nature of the resonant tunneling processes even at low doping concentrations.
This difference in behavior between the current and the gain with varying doping concentrations is attributed to the strong momentum dependence of the electron--charged-impurity interaction.

\begin{figure}
\begin{centering}
\includegraphics[width=0.235\textwidth]{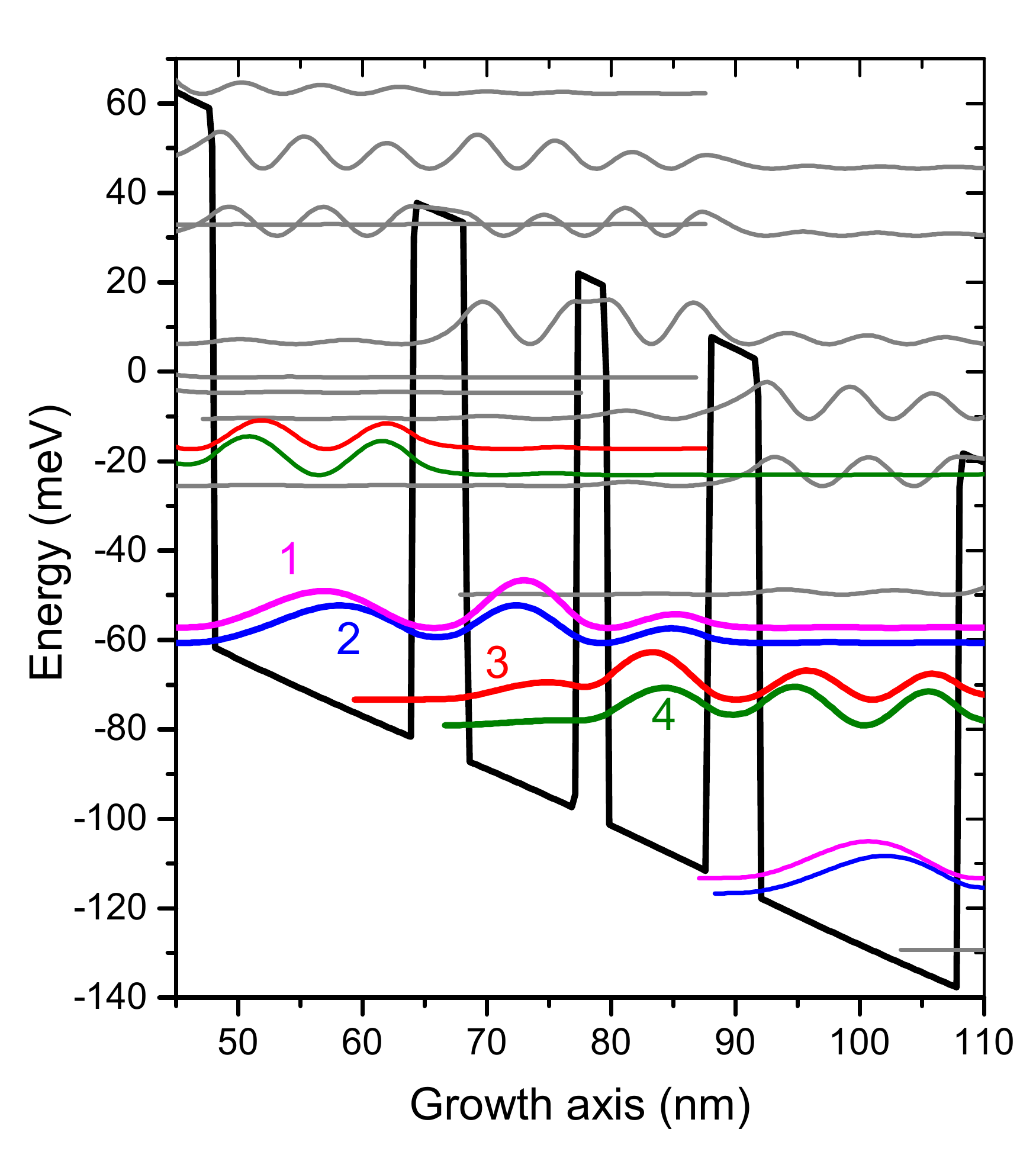} 
\includegraphics[width=0.235\textwidth]{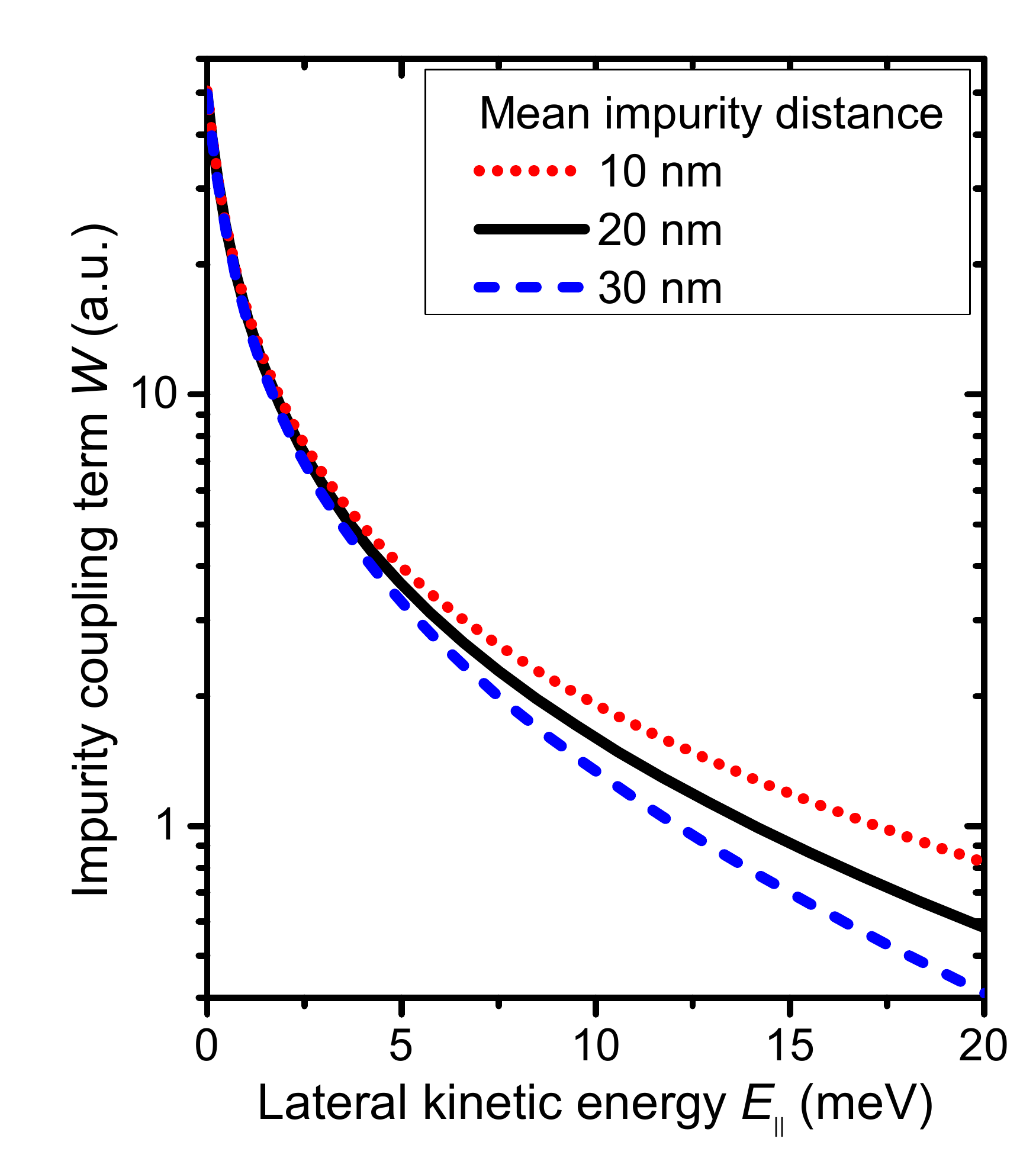} 
\end{centering}
\caption{(Color online). (a) Conduction band diagram of the studied THz QCL \cite{fathololoumi2012terahertz} for an electric field of 12.8~kV/cm.
The layer sequence of one period is $\bold{4.3}$/8.9/$\bold{2.46}$/8.15/$\bold{4.1}$/$\underline{16.0}$~nm with Al$_{0.15}$Ga$_{0.85}$As barriers in bold fonts and the doping region underlined (sheet doping density of $3\times10^{10}$cm$^{-2}$). A band offset of 120 meV is used for the GaAs/Al$_{0.15}$Ga$_{0.85}$As interface.
The Wannier--Stark levels considered in the calculation are represented, among which the lowest four levels labeled from 1 to 4.
(b) In-plane kinetic-energy dependence of the impurity interaction term $W$ for various mean axial distances to an ionized dopant.
}
\label{diagram}
\end{figure}

We consider a resonant-phonon THz QCL design similar to the one reported in Ref.~\onlinecite{fathololoumi2012terahertz}, in which a record operating temperature of almost 200K has been demonstrated. The conduction band diagram with the probability densities of the lowest Wannier--Stark states is shown in Fig.~\ref{diagram} at the designed operating bias. 
The NEGF calculations are performed in a cylindrical basis for the in-plane motion, extending the calculations reported in Refs.~\onlinecite{grange2014SLs} and \onlinecite{grange2014nanowire} for lateral confinement in nanowires. For large diameters, our calculations converge towards the two-dimensional limit for the in-plane motion \cite{grange2014SLs,grange2014nanowire}.
A mode-space basis is used in the axial direction, and field-periodic boundary conditions are used.
Inelastic scattering due to optical and acoustic phonons, as well as elastic scattering due to charged impurities, interface roughness and alloy disorder are taken into account within the self-consistent Born approximation.
Electron-electron interaction is treated within the mean-field approximation by solving self-consistently the corresponding Poisson equation.
The gain is calculated in a self-consistent way within the linear response theory \cite{wacker2002gain,banit2005self}.

The effect of electron-impurity interaction is separated into (i) a mean field which is found by solving self-consistently the Poisson equation for the electrostatic potential, and (ii) incoherent scattering processes (CIS) described by a self-energy.
To evaluate the latter, we express the screened Coulomb potential created by a given ionized dopant located at the position $\mathbf{r}_{\text{d}}$
\begin{equation}
V^{\text{(c)}}_{\mathbf{r}_{\text{d}}}(\mathbf{r}) = - \frac{e^2 }{4\pi \varepsilon_0 \varepsilon_{\text{s}} |\mathbf{r}-\mathbf{r}_{\text{d}}|} \exp \left( -\frac{|\mathbf{r}-\mathbf{r}_{\text{d}}|}{\lambda_{\text{s}}} \right) ,
\end{equation}
where $\varepsilon_0 \varepsilon_{\text{s}}$ is the static dielectric constant and $\lambda_{\text{s}}$ is the screening length. $\lambda_{\text{s}}$ is calculated within the Debye screening model for an equivalent 3D homogeneous electron gas with the same average electron density $\bar{n}_{\text{3D}}$:
\begin{equation}
\lambda_s = \sqrt{\frac{\epsilon_0 \varepsilon_{\text{s}}  k_{\text{B}} T_{\text{e}}}{e^2 \bar{n}_{\text{3D}}}} ,
\end{equation}
where $T_{\text{e}}$ is taken as an effective electron temperature. More precisely, we calculate the mean kinetic energy $\langle E_{\parallel} \rangle$ in the lateral directions. We then set the effective temperature $T_{\text{e}}$ so that $\langle E_{\parallel} \rangle$ matches the mean kinetic energy $\langle E_{\parallel}^{\text{eq}}(T_{\text{e}}) \rangle$ of a thermalized 2D electron gaz.
For a lattice temperature of 200~K studied in the following, $T_{\text{e}}$ is found to vary between 200~K and 234~K.
The above expression for the screened Coulomb potential is then used to evaluate the CIS self-energy within the self-consistent Born approximation \cite{grange2014SLs}.
It is instructive to express the coupling correlation between two axial positions ($z_1$,$z_2$) as a function of the scattered in-plane momentum $k_{\parallel}$ 
\begin{equation}
W(z_1,z_2,k_{\parallel}) \propto \frac{\exp \left( -2\overline{\Delta z}\sqrt{k_{\parallel}^2+1/\lambda_{\text{s}}^2} \right) }{k_{\parallel}^2+1/\lambda_{\text{s}}^2} ,
\end{equation}
where $\overline{\Delta z}=(|z_1-z_{\text{d}}|+|z_ 2-z_{\text{d}}|)/2$ is the mean axial distance to the ionized dopant. The momentum dependence of $W$ is plotted in Fig.~\ref{diagram} 	as a function of $E_{\parallel}=\hbar^2k_{\parallel}^2/2m^*$ for various typical mean axial distances to the impurities.
The importance of this strong momentum dependence is discussed below as the origin of the contrasting effects of CIS on transport and gain.
In contrast to previously reported NEGF studies of CIS, the full in-plane momentum dependence of the CIS is taken into account \cite{lee2002nonequilibrium,nelander2008temperature,nelander2009temperature}, and no approximation is made for the axial form factor \cite{kubis09}.


\begin{figure}
\begin{centering}
\includegraphics[width=0.48\textwidth]{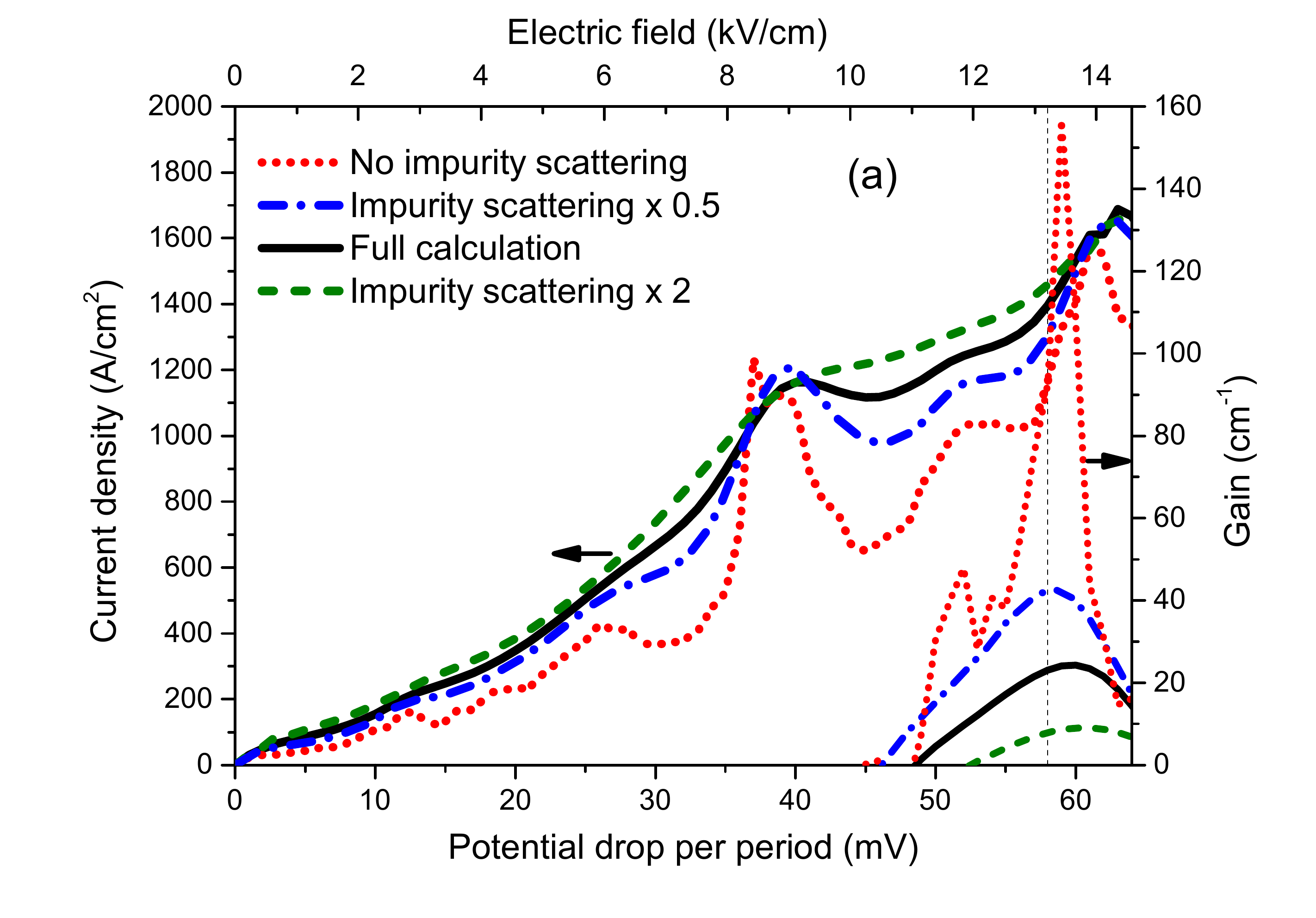} 
\includegraphics[width=0.48\textwidth]{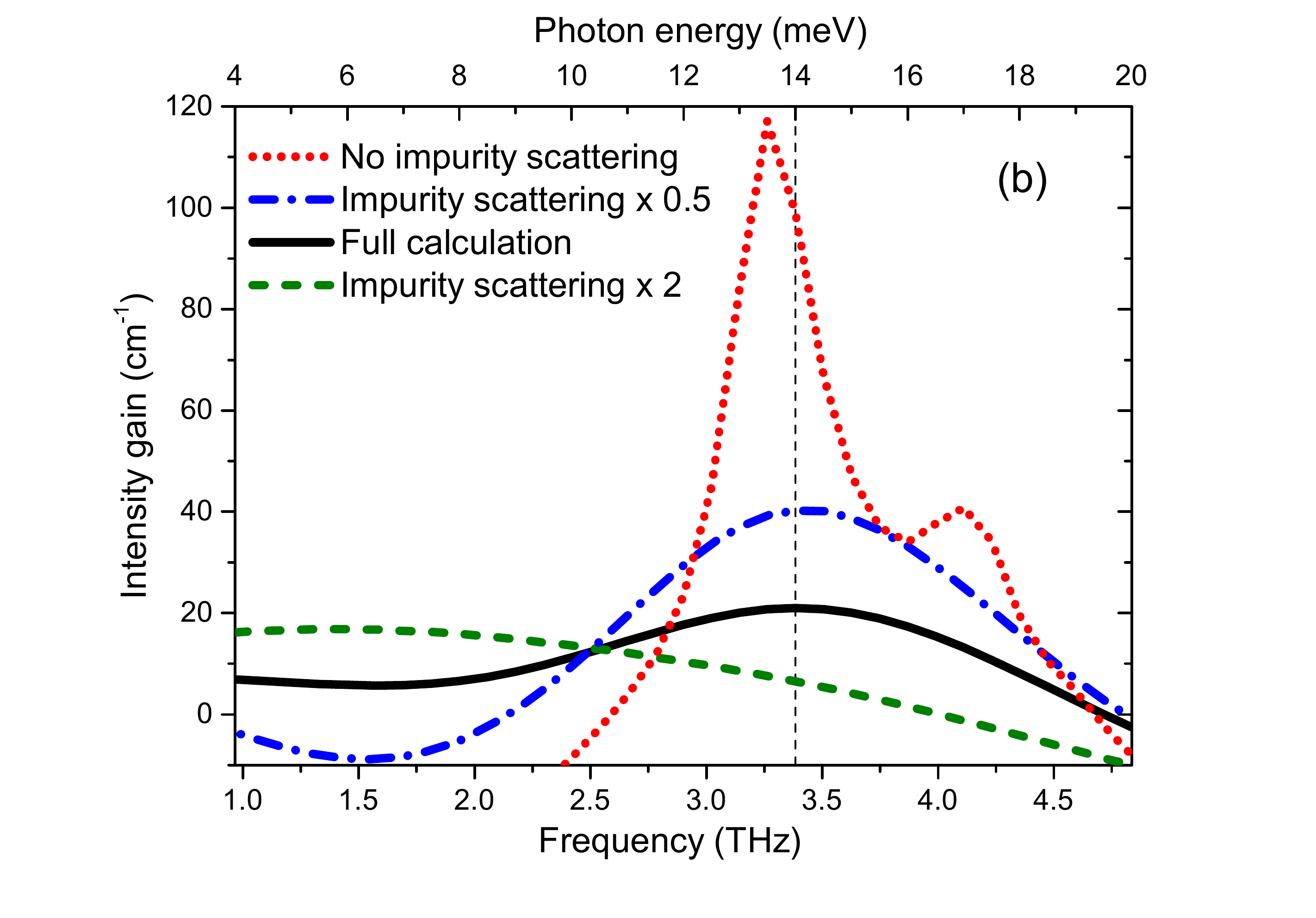} 
\end{centering}
\caption{(Color online). (a) Current-voltage characteristics (left scale) and intensity gain at 3.4~THz (right scale) as a function of voltage . The vertical dashed line indicates the voltage of 58 mV/period. (b) Intensity gain spectra for different strength of the charged impurity scattering at a voltage of 58 mV/period. The vertical dashed line indicates a photon energy of 14 meV. In (a) and (b) the sheet doping density is set to $3.10^{10}$cm$^{-2}$ and the temperature to 200K.}
\label{impur}
\end{figure}

Fig.~\ref{impur} displays current-voltage (I-V) characteristics (Fig.~\ref{impur}(a) left scale), the intensity gain at 3.4~THz versus voltage (Fig.~\ref{impur}(a) right scale) and the intensity gain spectrum (Fig.~\ref{impur}(b)) at a temperature of 200~K.
To describe the interface roughness, typical values of an amplitude of $0.1$~nm and an exponential correlation length of $8$~nm are assumed \cite{franckie2015impact}. Such interface roughness is found here to have a minor role with respect to the CIS discussed in the following.
The calculated I-V characteristics is consistent with the experimental one \cite{fathololoumi2012terahertz}, and the maximum gain at this maximum operating temperature of 200~K matches the typical losses measured in these THz cavities \cite{burghoff:261111}.
To investigate the sole effect of CIS, simulations in which only the self-energy describing CIS is artificially multiplied by a factor 0, 0.5, or 2 are also shown in Fig.~\ref{impur}. Mean-field effects are hence excluded and will be discussed later.
Without any impurity scattering, the shape of the I-V characteristics is remarkably different (Fig.~\ref{impur}(a)). It shows a marked resonance for a potential drop per period which matches the longitudinal-optical (LO) phonon energy (taken here as 36.7~meV).
For CIS strength being tuned from 0.5 to 2, more modest changes in the I-V characteristics are observed in Fig.~\ref{impur}(a).
In contrast, the gain properties are much more sensitive to these relative variations in the impurity scattering strength. For a CIS reduced by a factor 0.5, the maximum gain in Fig.~\ref{impur}(b) is almost twice (40 cm$^{-1}$ instead of 21cm$^{-1}$).

\begin{figure}
\begin{centering}
\includegraphics[width=0.48\textwidth]{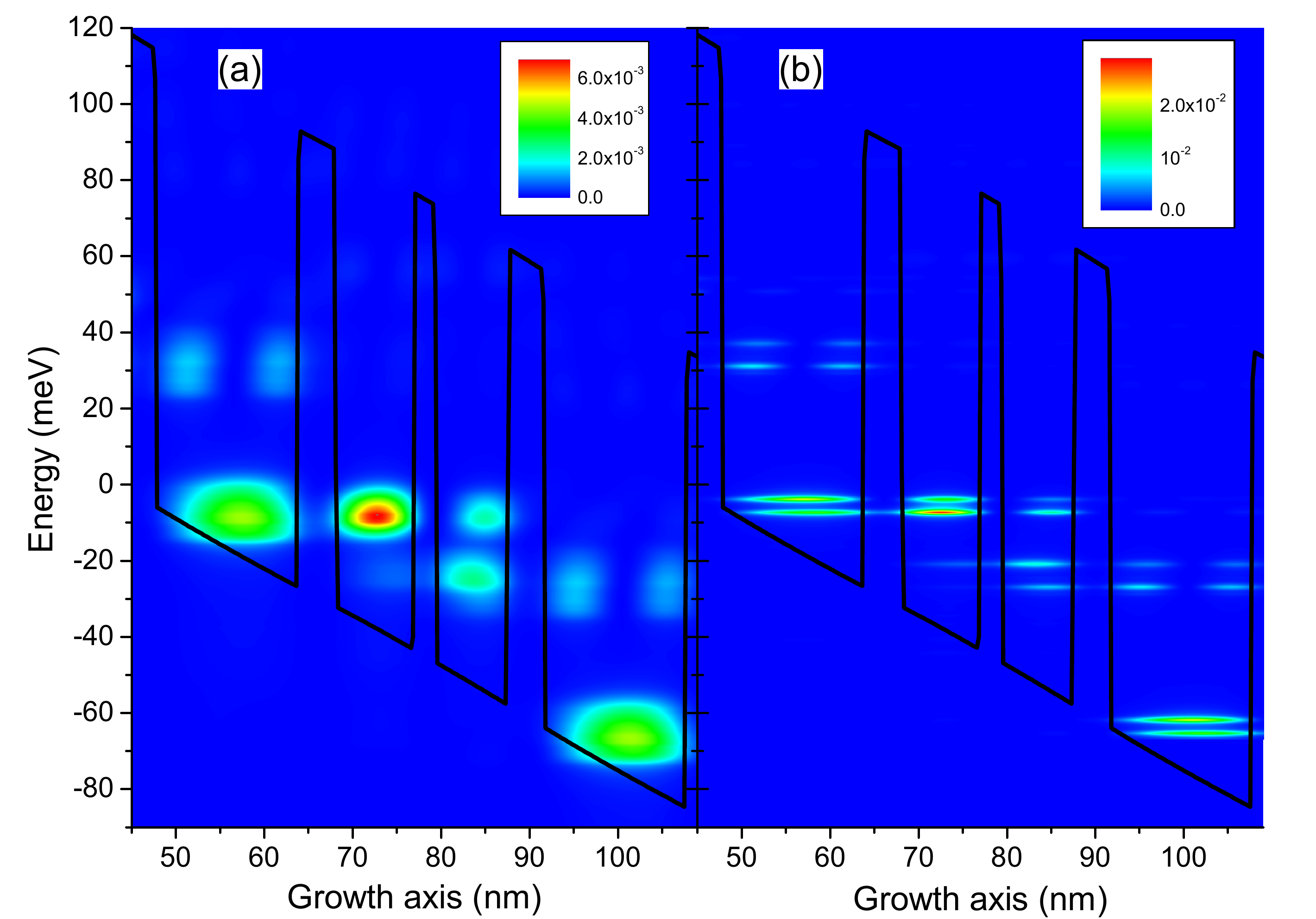}
\end{centering}
\caption{(Color online). Electron density $n(z,E )$ (arbitrary unit) resolved in energy and spatially along the growth axis for a voltage of 58~mV/period: (a) full calculation and (b) with impurity scattering turned off.}
\label{GLesser}
\end{figure}

To gain more insights into the role of CIS, the energy-resolved electron density is derived from the lesser Green's function and plotted in Fig.~\ref{GLesser} with and without considering CIS. In absence of CIS (Fig.~\ref{GLesser}b), the states across the thick tunnel barriers for injection and extraction clearly anticross, which indicates a coherent tunneling regime. In the full calculation (Fig.~\ref{GLesser}(a)), an incoherent tunneling regime is observed with much broader states.
As the impurity scattering is turned on, a transition from coherent to incoherent tunneling occurs when the impurity scattering strength reaches around 0.1 of its full value (i.e. corresponding to a doping density of $3\times10^9$~cm$^{-2}$ for the full calculation).

\begin{figure}
\begin{centering}
\includegraphics[width=0.5\textwidth]{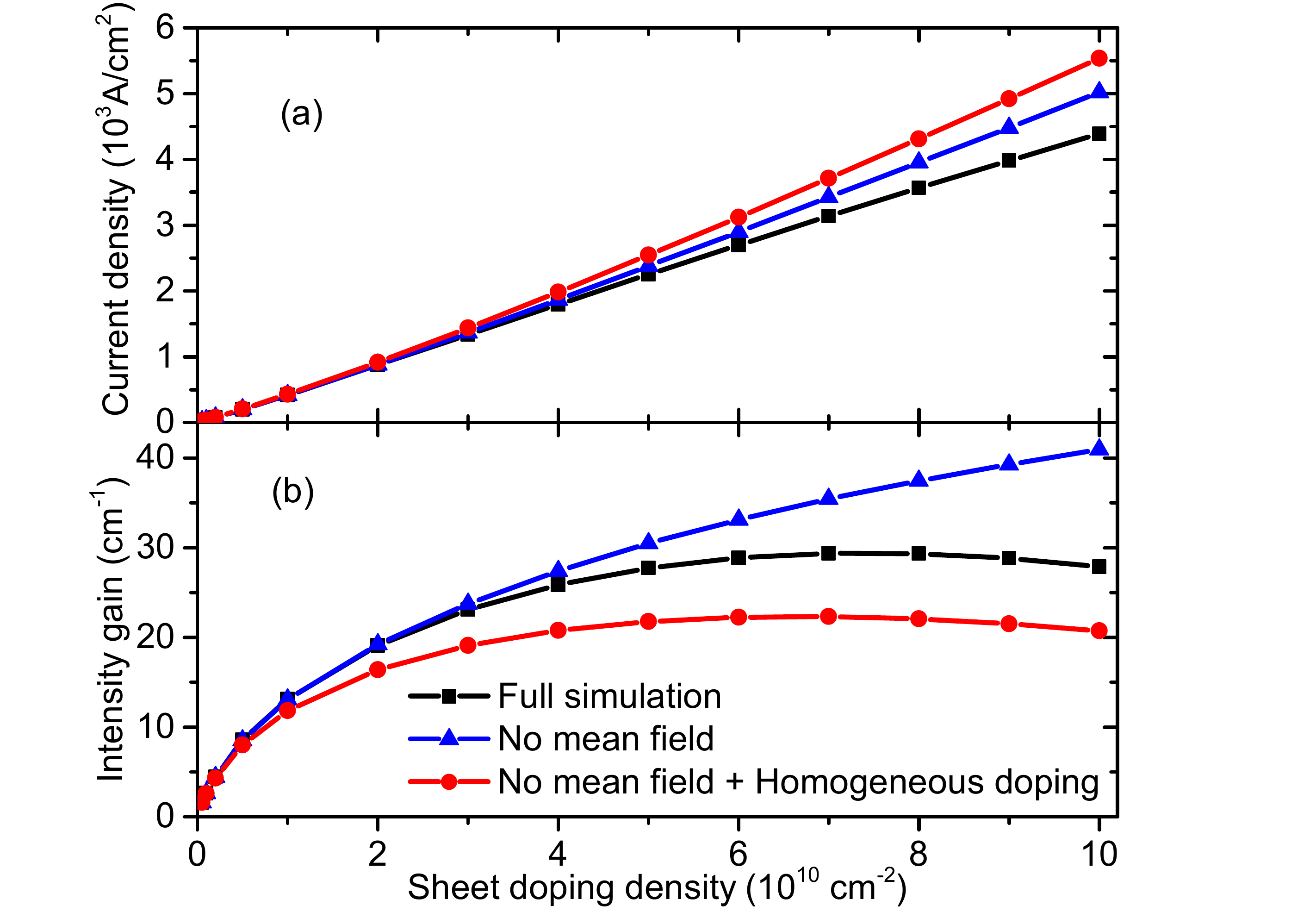}
\end{centering}
\caption{(Color online). (a) Current density and (b) intensity gain at 3.4 THz as the function of the sheet doping density per period. The full calculations are indicated by black squares;
calculations without considering mean field
 by blue triangles;
calculations without mean field and assuming an homogeneous doping density by red circles. }
\label{doping}
\end{figure}

Calculations of the doping dependence of the current density at the operating bias of 58mV/period and intensity gain at 3.4 THz are reported in Fig.~\ref{doping}.
Square symbols account for the full doping dependences, i.e. accounting for both CIS and mean-field effects.
The current density at the operating bias is found to evolve almost linearly with the doping density. In contrast the gain is found to saturate at the experimental doping density.
To distinguish between CIS and mean-field effects, calculations of the current and the gain assuming no mean-field interaction are also shown by triangles in Fig.~\ref{doping}.
The gain then shows a less pronounced saturation effect with increasing doping density, as the electric mean field starts to be significant for doping densities larger than $3 \times 10^{10}$~cm$^{-2}$.
The difference in the evolution of the gain between the full model and the no-mean-field model suggests that a further increase in the doping density should be accompanied by an adjustment of the design to compensate for the increasing mean field.
To investigate the influence on CIS of the spatial confinement of the impurities in the doping region, additional calculations assuming a homogeneous doping over the whole period and no mean field are indicated by solid circles. The saturation of the gain is then found to be much stronger than in the latter case, confirming that proximity of the dopants with the lasing states is detrimental.

We now discuss the origin of the contrasting effects of CIS on transport and gain.
As shown in Fig.~\ref{diagram}(b), the impurity interaction term is the largest for small momentum transfer, and rapidly decreases with increasing momentum.
Small-momentum scattering is expected to affect the spatial coherence of the electrons, but not to induce scattering among energetically distant levels like the laser states.
The large strength of small-momentum CIS explains why the transition from coherent to incoherent resonant tunneling occurs at relatively low doping concentration.
In addition, dephasing induced by small-momentum CIS broadens the gain line, which provides an important contribution to the decrease of the maximum gain as seen in Fig.~\ref{impur}(b). The fact that gain broadening was a major limitation of gain was already pointed out in Ref.~\onlinecite{nelander2008temperature}.
On the contrary, direct scattering among the laser states, involving a lateral kinetic energy around 14 meV, is much less favorable (see Fig.~\ref{diagram}(b)).
This explains why population inversion can still be obtained at the experimental doping concentration, whereas the transition from coherent to incoherent regime for the resonant tunneling processes occur one order of magnitude below.

To explain the relatively weak dependence of the current densities versus the doping dependence, in particular at the operating bias, one has to consider the opposite effects of CIS on resonant tunneling and on scattering among the laser states.
A decrease in the tunneling efficiency when going from coherent to incoherent tunneling regime is expected to occur. Yet, the current flow over the whole structure at the operating bias is mainly limited by the scattering from the upper to the lower laser level.
As the CIS between the laser levels is much less favorable, this explains why the change in the current is relatively modest with respect to the gain.

In summary, NEGF calculations including the full in-plane dependence of the charged impurity scattering are reported. A contrasting influence of impurity scattering on transport and gain is demonstrated, which is attributed to the long-range nature of the coupling to impurities. These new insights into the counter-intuitive role of impurity scattering suggest new routes for THz QCL optimization by engineering of the doping profile.


The author acknowledges support from the Alexander von Humboldt Foundation, the Austrian Science Fund FWF through Project No.~F25-P14 (SFB IR-ON) and the European Union Seventh Framework Programme (FP7) under Grant Agreement No.~618078 (WASPS).
S. Birner and B. Huber are acknowledged for careful reading of the manuscript.

\bibliographystyle{apsrev}
\bibliography{C:/Users/thomas.grange/SkyDrive/papers/biblio/biblio}

\begin{thebibliography}{40}
\expandafter\ifx\csname natexlab\endcsname\relax\def\natexlab#1{#1}\fi
\expandafter\ifx\csname bibnamefont\endcsname\relax
  \def\bibnamefont#1{#1}\fi
\expandafter\ifx\csname bibfnamefont\endcsname\relax
  \def\bibfnamefont#1{#1}\fi
\expandafter\ifx\csname citenamefont\endcsname\relax
  \def\citenamefont#1{#1}\fi
\expandafter\ifx\csname url\endcsname\relax
  \def\url#1{\texttt{#1}}\fi
\expandafter\ifx\csname urlprefix\endcsname\relax\def\urlprefix{URL }\fi
\providecommand{\bibinfo}[2]{#2}
\providecommand{\eprint}[2][]{\url{#2}}

\bibitem[{\citenamefont{K\"ohler et~al.}(2002)\citenamefont{K\"ohler,
  Tredicucci, Beltram, Beere, Linfield, Davies, Ritchie, Iotti, and
  Rossi}}]{Kohler02}
\bibinfo{author}{\bibfnamefont{R.}~\bibnamefont{K\"ohler}},
  \bibinfo{author}{\bibfnamefont{A.}~\bibnamefont{Tredicucci}},
  \bibinfo{author}{\bibfnamefont{F.}~\bibnamefont{Beltram}},
  \bibinfo{author}{\bibfnamefont{H.~E.} \bibnamefont{Beere}},
  \bibinfo{author}{\bibfnamefont{E.~H.} \bibnamefont{Linfield}},
  \bibinfo{author}{\bibfnamefont{A.~G.} \bibnamefont{Davies}},
  \bibinfo{author}{\bibfnamefont{D.~A.} \bibnamefont{Ritchie}},
  \bibinfo{author}{\bibfnamefont{R.~C.} \bibnamefont{Iotti}}, \bibnamefont{and}
  \bibinfo{author}{\bibfnamefont{F.}~\bibnamefont{Rossi}},
  \bibinfo{journal}{Nature} \textbf{\bibinfo{volume}{417}},
  \bibinfo{pages}{156} (\bibinfo{year}{2002}).

\bibitem[{\citenamefont{Williams}(2007)}]{williams07}
\bibinfo{author}{\bibfnamefont{B.~S.} \bibnamefont{Williams}},
  \bibinfo{journal}{Nat. Photonics} \textbf{\bibinfo{volume}{1}},
  \bibinfo{pages}{517} (\bibinfo{year}{2007}).

\bibitem[{\citenamefont{Vitiello et~al.}(2015)\citenamefont{Vitiello, Scalari,
  Williams, and De~Natale}}]{vitiello2015quantum}
\bibinfo{author}{\bibfnamefont{M.~S.} \bibnamefont{Vitiello}},
  \bibinfo{author}{\bibfnamefont{G.}~\bibnamefont{Scalari}},
  \bibinfo{author}{\bibfnamefont{B.}~\bibnamefont{Williams}}, \bibnamefont{and}
  \bibinfo{author}{\bibfnamefont{P.}~\bibnamefont{De~Natale}},
  \bibinfo{journal}{Opt. Express} \textbf{\bibinfo{volume}{23}},
  \bibinfo{pages}{5167} (\bibinfo{year}{2015}).

\bibitem[{\citenamefont{Li et~al.}(2014)\citenamefont{Li, Chen, Zhu, Freeman,
  Dean, Valavanis, Linfield et~al.}}]{li2014terahertz}
\bibinfo{author}{\bibfnamefont{L.}~\bibnamefont{Li}},
  \bibinfo{author}{\bibfnamefont{L.}~\bibnamefont{Chen}},
  \bibinfo{author}{\bibfnamefont{J.}~\bibnamefont{Zhu}},
  \bibinfo{author}{\bibfnamefont{J.}~\bibnamefont{Freeman}},
  \bibinfo{author}{\bibfnamefont{P.}~\bibnamefont{Dean}},
  \bibinfo{author}{\bibfnamefont{A.}~\bibnamefont{Valavanis}},
  \bibinfo{author}{\bibfnamefont{E.~H.} \bibnamefont{Linfield}},
  \bibnamefont{et~al.}, \bibinfo{journal}{Electron. Lett.}
  \textbf{\bibinfo{volume}{50}}, \bibinfo{pages}{309} (\bibinfo{year}{2014}).

\bibitem[{\citenamefont{Liu et~al.}(2005)\citenamefont{Liu, W{\"a}chter, Ban,
  Wasilewski, Buchanan, Aers, Cao, Feng, Williams, and Hu}}]{liu2005effect}
\bibinfo{author}{\bibfnamefont{H.}~\bibnamefont{Liu}},
  \bibinfo{author}{\bibfnamefont{M.}~\bibnamefont{W{\"a}chter}},
  \bibinfo{author}{\bibfnamefont{D.}~\bibnamefont{Ban}},
  \bibinfo{author}{\bibfnamefont{Z.}~\bibnamefont{Wasilewski}},
  \bibinfo{author}{\bibfnamefont{M.}~\bibnamefont{Buchanan}},
  \bibinfo{author}{\bibfnamefont{G.}~\bibnamefont{Aers}},
  \bibinfo{author}{\bibfnamefont{J.}~\bibnamefont{Cao}},
  \bibinfo{author}{\bibfnamefont{S.}~\bibnamefont{Feng}},
  \bibinfo{author}{\bibfnamefont{B.}~\bibnamefont{Williams}}, \bibnamefont{and}
  \bibinfo{author}{\bibfnamefont{Q.}~\bibnamefont{Hu}}, \bibinfo{journal}{Appl.
  Phys. Lett.} \textbf{\bibinfo{volume}{87}}, \bibinfo{pages}{141102}
  (\bibinfo{year}{2005}).

\bibitem[{\citenamefont{Ajili et~al.}(2006)\citenamefont{Ajili, Scalari,
  Giovannini, Hoyler, and Faist}}]{ajili2006doping}
\bibinfo{author}{\bibfnamefont{L.}~\bibnamefont{Ajili}},
  \bibinfo{author}{\bibfnamefont{G.}~\bibnamefont{Scalari}},
  \bibinfo{author}{\bibfnamefont{M.}~\bibnamefont{Giovannini}},
  \bibinfo{author}{\bibfnamefont{N.}~\bibnamefont{Hoyler}}, \bibnamefont{and}
  \bibinfo{author}{\bibfnamefont{J.}~\bibnamefont{Faist}}, \bibinfo{journal}{J.
  Appl. Phys.} \textbf{\bibinfo{volume}{100}}, \bibinfo{pages}{043102}
  (\bibinfo{year}{2006}).

\bibitem[{\citenamefont{Benz et~al.}(2007)\citenamefont{Benz, Fasching,
  Andrews, Martl, Unterrainer, Roch, Schrenk, Golka, and
  Strasser}}]{benz2007influence}
\bibinfo{author}{\bibfnamefont{A.}~\bibnamefont{Benz}},
  \bibinfo{author}{\bibfnamefont{G.}~\bibnamefont{Fasching}},
  \bibinfo{author}{\bibfnamefont{A.~M.} \bibnamefont{Andrews}},
  \bibinfo{author}{\bibfnamefont{M.}~\bibnamefont{Martl}},
  \bibinfo{author}{\bibfnamefont{K.}~\bibnamefont{Unterrainer}},
  \bibinfo{author}{\bibfnamefont{T.}~\bibnamefont{Roch}},
  \bibinfo{author}{\bibfnamefont{W.}~\bibnamefont{Schrenk}},
  \bibinfo{author}{\bibfnamefont{S.}~\bibnamefont{Golka}}, \bibnamefont{and}
  \bibinfo{author}{\bibfnamefont{G.}~\bibnamefont{Strasser}},
  \bibinfo{journal}{Appl. Phys. Lett.} \textbf{\bibinfo{volume}{90}}
  (\bibinfo{year}{2007}).

\bibitem[{\citenamefont{Andrews et~al.}(2008)\citenamefont{Andrews, Benz,
  Deutsch, Fasching, Unterrainer, Klang, Schrenk, and
  Strasser}}]{andrews2008doping}
\bibinfo{author}{\bibfnamefont{A.~M.} \bibnamefont{Andrews}},
  \bibinfo{author}{\bibfnamefont{A.}~\bibnamefont{Benz}},
  \bibinfo{author}{\bibfnamefont{C.}~\bibnamefont{Deutsch}},
  \bibinfo{author}{\bibfnamefont{G.}~\bibnamefont{Fasching}},
  \bibinfo{author}{\bibfnamefont{K.}~\bibnamefont{Unterrainer}},
  \bibinfo{author}{\bibfnamefont{P.}~\bibnamefont{Klang}},
  \bibinfo{author}{\bibfnamefont{W.}~\bibnamefont{Schrenk}}, \bibnamefont{and}
  \bibinfo{author}{\bibfnamefont{G.}~\bibnamefont{Strasser}},
  \bibinfo{journal}{Mater. Sci. Eng. B} \textbf{\bibinfo{volume}{147}},
  \bibinfo{pages}{152} (\bibinfo{year}{2008}).

\bibitem[{\citenamefont{Deutsch et~al.}(2013)\citenamefont{Deutsch, Detz,
  Krall, Brandstetter, Zederbauer, Andrews, Schrenk, Strasser, and
  Unterrainer}}]{deutsch2013dopant}
\bibinfo{author}{\bibfnamefont{C.}~\bibnamefont{Deutsch}},
  \bibinfo{author}{\bibfnamefont{H.}~\bibnamefont{Detz}},
  \bibinfo{author}{\bibfnamefont{M.}~\bibnamefont{Krall}},
  \bibinfo{author}{\bibfnamefont{M.}~\bibnamefont{Brandstetter}},
  \bibinfo{author}{\bibfnamefont{T.}~\bibnamefont{Zederbauer}},
  \bibinfo{author}{\bibfnamefont{A.}~\bibnamefont{Andrews}},
  \bibinfo{author}{\bibfnamefont{W.}~\bibnamefont{Schrenk}},
  \bibinfo{author}{\bibfnamefont{G.}~\bibnamefont{Strasser}}, \bibnamefont{and}
  \bibinfo{author}{\bibfnamefont{K.}~\bibnamefont{Unterrainer}},
  \bibinfo{journal}{Appl. Phys. Lett.} \textbf{\bibinfo{volume}{102}},
  \bibinfo{pages}{201102} (\bibinfo{year}{2013}).

\bibitem[{\citenamefont{Callebaut et~al.}(2004)\citenamefont{Callebaut, Kumar,
  Williams, Hu, and Reno}}]{callebaut2004importance}
\bibinfo{author}{\bibfnamefont{H.}~\bibnamefont{Callebaut}},
  \bibinfo{author}{\bibfnamefont{S.}~\bibnamefont{Kumar}},
  \bibinfo{author}{\bibfnamefont{B.~S.} \bibnamefont{Williams}},
  \bibinfo{author}{\bibfnamefont{Q.}~\bibnamefont{Hu}}, \bibnamefont{and}
  \bibinfo{author}{\bibfnamefont{J.~L.} \bibnamefont{Reno}},
  \bibinfo{journal}{Appl. Phys. Lett.} \textbf{\bibinfo{volume}{84}},
  \bibinfo{pages}{645} (\bibinfo{year}{2004}).

\bibitem[{\citenamefont{Banit et~al.}(2005)\citenamefont{Banit, Lee, Knorr, and
  Wacker}}]{banit2005self}
\bibinfo{author}{\bibfnamefont{F.}~\bibnamefont{Banit}},
  \bibinfo{author}{\bibfnamefont{S.}~\bibnamefont{Lee}},
  \bibinfo{author}{\bibfnamefont{A.}~\bibnamefont{Knorr}}, \bibnamefont{and}
  \bibinfo{author}{\bibfnamefont{A.}~\bibnamefont{Wacker}},
  \bibinfo{journal}{Appl. Phys. Lett.} \textbf{\bibinfo{volume}{86}},
  \bibinfo{pages}{041108} (\bibinfo{year}{2005}).

\bibitem[{\citenamefont{Nelander and Wacker}(2008)}]{nelander2008temperature}
\bibinfo{author}{\bibfnamefont{R.}~\bibnamefont{Nelander}} \bibnamefont{and}
  \bibinfo{author}{\bibfnamefont{A.}~\bibnamefont{Wacker}},
  \bibinfo{journal}{Appl. Phys. Lett.} \textbf{\bibinfo{volume}{92}},
  \bibinfo{pages}{081102} (\bibinfo{year}{2008}).

\bibitem[{\citenamefont{Nelander and Wacker}(2009)}]{nelander2009temperature}
\bibinfo{author}{\bibfnamefont{R.}~\bibnamefont{Nelander}} \bibnamefont{and}
  \bibinfo{author}{\bibfnamefont{A.}~\bibnamefont{Wacker}},
  \bibinfo{journal}{J. Appl. Phys.} \textbf{\bibinfo{volume}{106}},
  \bibinfo{pages}{063115} (\bibinfo{year}{2009}).

\bibitem[{\citenamefont{Pereira~Jr et~al.}(2004)\citenamefont{Pereira~Jr, Lee,
  and Wacker}}]{pereira2004controlling}
\bibinfo{author}{\bibfnamefont{M.}~\bibnamefont{Pereira~Jr}},
  \bibinfo{author}{\bibfnamefont{S.-C.} \bibnamefont{Lee}}, \bibnamefont{and}
  \bibinfo{author}{\bibfnamefont{A.}~\bibnamefont{Wacker}},
  \bibinfo{journal}{Phys. Rev. B} \textbf{\bibinfo{volume}{69}},
  \bibinfo{pages}{205310} (\bibinfo{year}{2004}).

\bibitem[{\citenamefont{Jovanovi{\'c} et~al.}(2005)\citenamefont{Jovanovi{\'c},
  Indjin, Vukmirovi{\'c}, Ikoni{\'c}, Harrison, Linfield, Page, Marcadet,
  Sirtori, Worrall et~al.}}]{jovanovic2005mechanisms}
\bibinfo{author}{\bibfnamefont{V.}~\bibnamefont{Jovanovi{\'c}}},
  \bibinfo{author}{\bibfnamefont{D.}~\bibnamefont{Indjin}},
  \bibinfo{author}{\bibfnamefont{N.}~\bibnamefont{Vukmirovi{\'c}}},
  \bibinfo{author}{\bibfnamefont{Z.}~\bibnamefont{Ikoni{\'c}}},
  \bibinfo{author}{\bibfnamefont{P.}~\bibnamefont{Harrison}},
  \bibinfo{author}{\bibfnamefont{E.}~\bibnamefont{Linfield}},
  \bibinfo{author}{\bibfnamefont{H.}~\bibnamefont{Page}},
  \bibinfo{author}{\bibfnamefont{X.}~\bibnamefont{Marcadet}},
  \bibinfo{author}{\bibfnamefont{C.}~\bibnamefont{Sirtori}},
  \bibinfo{author}{\bibfnamefont{C.}~\bibnamefont{Worrall}},
  \bibnamefont{et~al.}, \bibinfo{journal}{Appl. Phys. Lett.}
  \textbf{\bibinfo{volume}{86}}, \bibinfo{pages}{211117}
  (\bibinfo{year}{2005}).

\bibitem[{\citenamefont{Leuliet et~al.}(2006)\citenamefont{Leuliet, Vasanelli,
  Wade, Fedorov, Smirnov, Bastard, and Sirtori}}]{leuliet2006electron}
\bibinfo{author}{\bibfnamefont{A.}~\bibnamefont{Leuliet}},
  \bibinfo{author}{\bibfnamefont{A.}~\bibnamefont{Vasanelli}},
  \bibinfo{author}{\bibfnamefont{A.}~\bibnamefont{Wade}},
  \bibinfo{author}{\bibfnamefont{G.}~\bibnamefont{Fedorov}},
  \bibinfo{author}{\bibfnamefont{D.}~\bibnamefont{Smirnov}},
  \bibinfo{author}{\bibfnamefont{G.}~\bibnamefont{Bastard}}, \bibnamefont{and}
  \bibinfo{author}{\bibfnamefont{C.}~\bibnamefont{Sirtori}},
  \bibinfo{journal}{Phys. Rev. B} \textbf{\bibinfo{volume}{73}},
  \bibinfo{pages}{085311} (\bibinfo{year}{2006}).

\bibitem[{\citenamefont{Jirauschek et~al.}(2007)\citenamefont{Jirauschek,
  Scarpa, Lugli, Vitiello, and Scamarcio}}]{jirauschek2007comparative}
\bibinfo{author}{\bibfnamefont{C.}~\bibnamefont{Jirauschek}},
  \bibinfo{author}{\bibfnamefont{G.}~\bibnamefont{Scarpa}},
  \bibinfo{author}{\bibfnamefont{P.}~\bibnamefont{Lugli}},
  \bibinfo{author}{\bibfnamefont{M.~S.} \bibnamefont{Vitiello}},
  \bibnamefont{and}
  \bibinfo{author}{\bibfnamefont{G.}~\bibnamefont{Scamarcio}},
  \bibinfo{journal}{J. Appl. Phys.} \textbf{\bibinfo{volume}{101}},
  \bibinfo{pages}{086109} (\bibinfo{year}{2007}).

\bibitem[{\citenamefont{Jirauschek and Lugli}(2009)}]{jirauschek2009monte}
\bibinfo{author}{\bibfnamefont{C.}~\bibnamefont{Jirauschek}} \bibnamefont{and}
  \bibinfo{author}{\bibfnamefont{P.}~\bibnamefont{Lugli}}, \bibinfo{journal}{J.
  Appl. Phys.} \textbf{\bibinfo{volume}{105}}, \bibinfo{pages}{123102}
  (\bibinfo{year}{2009}).

\bibitem[{\citenamefont{Kumar and Hu}(2009)}]{Kumar09b}
\bibinfo{author}{\bibfnamefont{S.}~\bibnamefont{Kumar}} \bibnamefont{and}
  \bibinfo{author}{\bibfnamefont{Q.}~\bibnamefont{Hu}}, \bibinfo{journal}{Phys.
  Rev. B} \textbf{\bibinfo{volume}{80}}, \bibinfo{pages}{245316}
  (\bibinfo{year}{2009}).

\bibitem[{\citenamefont{Terazzi and Faist}(2010)}]{terazzi2010density}
\bibinfo{author}{\bibfnamefont{R.}~\bibnamefont{Terazzi}} \bibnamefont{and}
  \bibinfo{author}{\bibfnamefont{J.}~\bibnamefont{Faist}},
  \bibinfo{journal}{New J. Phys.} \textbf{\bibinfo{volume}{12}},
  \bibinfo{pages}{033045} (\bibinfo{year}{2010}).

\bibitem[{\citenamefont{Dupont et~al.}(2010)\citenamefont{Dupont, Fathololoumi,
  and Liu}}]{dupont2010simplified}
\bibinfo{author}{\bibfnamefont{E.}~\bibnamefont{Dupont}},
  \bibinfo{author}{\bibfnamefont{S.}~\bibnamefont{Fathololoumi}},
  \bibnamefont{and} \bibinfo{author}{\bibfnamefont{H.}~\bibnamefont{Liu}},
  \bibinfo{journal}{Phys. Rev. B} \textbf{\bibinfo{volume}{81}},
  \bibinfo{pages}{205311} (\bibinfo{year}{2010}).

\bibitem[{\citenamefont{Wacker et~al.}(2011)\citenamefont{Wacker, Bastard,
  Carosella, Ferreira, and Dupont}}]{wacker2011unraveling}
\bibinfo{author}{\bibfnamefont{A.}~\bibnamefont{Wacker}},
  \bibinfo{author}{\bibfnamefont{G.}~\bibnamefont{Bastard}},
  \bibinfo{author}{\bibfnamefont{F.}~\bibnamefont{Carosella}},
  \bibinfo{author}{\bibfnamefont{R.}~\bibnamefont{Ferreira}}, \bibnamefont{and}
  \bibinfo{author}{\bibfnamefont{E.}~\bibnamefont{Dupont}},
  \bibinfo{journal}{Phys. Rev. B} \textbf{\bibinfo{volume}{84}},
  \bibinfo{pages}{205319} (\bibinfo{year}{2011}).

\bibitem[{\citenamefont{Carosella et~al.}(2012)\citenamefont{Carosella,
  Ndebeka-Bandou, Ferreira, Dupont, Unterrainer, Strasser, Wacker, and
  Bastard}}]{carosella2012free}
\bibinfo{author}{\bibfnamefont{F.}~\bibnamefont{Carosella}},
  \bibinfo{author}{\bibfnamefont{C.}~\bibnamefont{Ndebeka-Bandou}},
  \bibinfo{author}{\bibfnamefont{R.}~\bibnamefont{Ferreira}},
  \bibinfo{author}{\bibfnamefont{E.}~\bibnamefont{Dupont}},
  \bibinfo{author}{\bibfnamefont{K.}~\bibnamefont{Unterrainer}},
  \bibinfo{author}{\bibfnamefont{G.}~\bibnamefont{Strasser}},
  \bibinfo{author}{\bibfnamefont{A.}~\bibnamefont{Wacker}}, \bibnamefont{and}
  \bibinfo{author}{\bibfnamefont{G.}~\bibnamefont{Bastard}},
  \bibinfo{journal}{Phys. Rev. B} \textbf{\bibinfo{volume}{85}},
  \bibinfo{pages}{085310} (\bibinfo{year}{2012}).

\bibitem[{\citenamefont{Matyas et~al.}(2013)\citenamefont{Matyas, Lugli, and
  Jirauschek}}]{matyas2013role}
\bibinfo{author}{\bibfnamefont{A.}~\bibnamefont{Matyas}},
  \bibinfo{author}{\bibfnamefont{P.}~\bibnamefont{Lugli}}, \bibnamefont{and}
  \bibinfo{author}{\bibfnamefont{C.}~\bibnamefont{Jirauschek}},
  \bibinfo{journal}{Appl. Phys. Lett.} \textbf{\bibinfo{volume}{102}},
  \bibinfo{pages}{011101} (\bibinfo{year}{2013}).

\bibitem[{\citenamefont{Ndebeka-Bandou
  et~al.}(2013)\citenamefont{Ndebeka-Bandou, Wacker, Carosella, Ferreira, and
  Bastard}}]{ndebeka2013dopant}
\bibinfo{author}{\bibfnamefont{C.}~\bibnamefont{Ndebeka-Bandou}},
  \bibinfo{author}{\bibfnamefont{A.}~\bibnamefont{Wacker}},
  \bibinfo{author}{\bibfnamefont{F.}~\bibnamefont{Carosella}},
  \bibinfo{author}{\bibfnamefont{R.}~\bibnamefont{Ferreira}}, \bibnamefont{and}
  \bibinfo{author}{\bibfnamefont{G.}~\bibnamefont{Bastard}},
  \bibinfo{journal}{Appl. Phys. Express} \textbf{\bibinfo{volume}{6}},
  \bibinfo{pages}{094101} (\bibinfo{year}{2013}).

\bibitem[{\citenamefont{Jirauschek and Kubis}(2014)}]{jirauschek2014modeling}
\bibinfo{author}{\bibfnamefont{C.}~\bibnamefont{Jirauschek}} \bibnamefont{and}
  \bibinfo{author}{\bibfnamefont{T.}~\bibnamefont{Kubis}},
  \bibinfo{journal}{Appl. Phys. Rev.} \textbf{\bibinfo{volume}{1}},
  \bibinfo{pages}{011307} (\bibinfo{year}{2014}).

\bibitem[{\citenamefont{Agnew et~al.}(2015)\citenamefont{Agnew, Grier, Taimre,
  Lim, Nikoli{\'c}, Valavanis, Cooper, Dean, Khanna, Lachab
  et~al.}}]{agnew2015efficient}
\bibinfo{author}{\bibfnamefont{G.}~\bibnamefont{Agnew}},
  \bibinfo{author}{\bibfnamefont{A.}~\bibnamefont{Grier}},
  \bibinfo{author}{\bibfnamefont{T.}~\bibnamefont{Taimre}},
  \bibinfo{author}{\bibfnamefont{Y.}~\bibnamefont{Lim}},
  \bibinfo{author}{\bibfnamefont{M.}~\bibnamefont{Nikoli{\'c}}},
  \bibinfo{author}{\bibfnamefont{A.}~\bibnamefont{Valavanis}},
  \bibinfo{author}{\bibfnamefont{J.}~\bibnamefont{Cooper}},
  \bibinfo{author}{\bibfnamefont{P.}~\bibnamefont{Dean}},
  \bibinfo{author}{\bibfnamefont{S.}~\bibnamefont{Khanna}},
  \bibinfo{author}{\bibfnamefont{M.}~\bibnamefont{Lachab}},
  \bibnamefont{et~al.}, \bibinfo{journal}{Appl. Phys. Lett.}
  \textbf{\bibinfo{volume}{106}}, \bibinfo{pages}{161105}
  (\bibinfo{year}{2015}).

\bibitem[{\citenamefont{Francki{\'e} et~al.}(2015)\citenamefont{Francki{\'e},
  Winge, Wolf, Liverini, Dupont, Trinit{\'e}, Faist, and
  Wacker}}]{franckie2015impact}
\bibinfo{author}{\bibfnamefont{M.}~\bibnamefont{Francki{\'e}}},
  \bibinfo{author}{\bibfnamefont{D.~O.} \bibnamefont{Winge}},
  \bibinfo{author}{\bibfnamefont{J.}~\bibnamefont{Wolf}},
  \bibinfo{author}{\bibfnamefont{V.}~\bibnamefont{Liverini}},
  \bibinfo{author}{\bibfnamefont{E.}~\bibnamefont{Dupont}},
  \bibinfo{author}{\bibfnamefont{V.}~\bibnamefont{Trinit{\'e}}},
  \bibinfo{author}{\bibfnamefont{J.}~\bibnamefont{Faist}}, \bibnamefont{and}
  \bibinfo{author}{\bibfnamefont{A.}~\bibnamefont{Wacker}},
  \bibinfo{journal}{Opt. Express} \textbf{\bibinfo{volume}{23}},
  \bibinfo{pages}{5201} (\bibinfo{year}{2015}).

\bibitem[{\citenamefont{Lee and Wacker}(2002)}]{lee2002nonequilibrium}
\bibinfo{author}{\bibfnamefont{S.}~\bibnamefont{Lee}} \bibnamefont{and}
  \bibinfo{author}{\bibfnamefont{A.}~\bibnamefont{Wacker}},
  \bibinfo{journal}{Phys. Rev. B} \textbf{\bibinfo{volume}{66}},
  \bibinfo{pages}{245314} (\bibinfo{year}{2002}).

\bibitem[{\citenamefont{Wacker}(2002)}]{wacker2002gain}
\bibinfo{author}{\bibfnamefont{A.}~\bibnamefont{Wacker}},
  \bibinfo{journal}{Phys. Rev. B} \textbf{\bibinfo{volume}{66}},
  \bibinfo{pages}{085326} (\bibinfo{year}{2002}).

\bibitem[{\citenamefont{Lee et~al.}(2006)\citenamefont{Lee, Banit, Woerner, and
  Wacker}}]{lee2006quantum}
\bibinfo{author}{\bibfnamefont{S.}~\bibnamefont{Lee}},
  \bibinfo{author}{\bibfnamefont{F.}~\bibnamefont{Banit}},
  \bibinfo{author}{\bibfnamefont{M.}~\bibnamefont{Woerner}}, \bibnamefont{and}
  \bibinfo{author}{\bibfnamefont{A.}~\bibnamefont{Wacker}},
  \bibinfo{journal}{Phys. Rev. B} \textbf{\bibinfo{volume}{73}},
  \bibinfo{pages}{245320} (\bibinfo{year}{2006}).

\bibitem[{\citenamefont{Kubis et~al.}(2009)\citenamefont{Kubis, Yeh, Vogl,
  Benz, Fasching, and Deutsch}}]{kubis09}
\bibinfo{author}{\bibfnamefont{T.}~\bibnamefont{Kubis}},
  \bibinfo{author}{\bibfnamefont{C.}~\bibnamefont{Yeh}},
  \bibinfo{author}{\bibfnamefont{P.}~\bibnamefont{Vogl}},
  \bibinfo{author}{\bibfnamefont{A.}~\bibnamefont{Benz}},
  \bibinfo{author}{\bibfnamefont{G.}~\bibnamefont{Fasching}}, \bibnamefont{and}
  \bibinfo{author}{\bibfnamefont{C.}~\bibnamefont{Deutsch}},
  \bibinfo{journal}{Phys. Rev. B} \textbf{\bibinfo{volume}{79}},
  \bibinfo{pages}{195323} (\bibinfo{year}{2009}).

\bibitem[{\citenamefont{Schmielau and
  Pereira~Jr}(2009)}]{schmielau2009nonequilibrium}
\bibinfo{author}{\bibfnamefont{T.}~\bibnamefont{Schmielau}} \bibnamefont{and}
  \bibinfo{author}{\bibfnamefont{M.}~\bibnamefont{Pereira~Jr}},
  \bibinfo{journal}{Appl. Phys. Lett.} \textbf{\bibinfo{volume}{95}},
  \bibinfo{pages}{231111} (\bibinfo{year}{2009}).

\bibitem[{\citenamefont{Yasuda et~al.}(2009)\citenamefont{Yasuda, Kubis, Vogl,
  Sekine, Hosako, and Hirakawa}}]{yasuda2009nonequilibrium}
\bibinfo{author}{\bibfnamefont{H.}~\bibnamefont{Yasuda}},
  \bibinfo{author}{\bibfnamefont{T.}~\bibnamefont{Kubis}},
  \bibinfo{author}{\bibfnamefont{P.}~\bibnamefont{Vogl}},
  \bibinfo{author}{\bibfnamefont{N.}~\bibnamefont{Sekine}},
  \bibinfo{author}{\bibfnamefont{I.}~\bibnamefont{Hosako}}, \bibnamefont{and}
  \bibinfo{author}{\bibfnamefont{K.}~\bibnamefont{Hirakawa}},
  \bibinfo{journal}{Appl. Phys. Lett.} \textbf{\bibinfo{volume}{94}},
  \bibinfo{pages}{151109} (\bibinfo{year}{2009}).

\bibitem[{\citenamefont{Kubis et~al.}(2010)\citenamefont{Kubis, Mehrotra, and
  Klimeck}}]{kubis2010design}
\bibinfo{author}{\bibfnamefont{T.}~\bibnamefont{Kubis}},
  \bibinfo{author}{\bibfnamefont{S.~R.} \bibnamefont{Mehrotra}},
  \bibnamefont{and} \bibinfo{author}{\bibfnamefont{G.}~\bibnamefont{Klimeck}},
  \bibinfo{journal}{Appl. Phys. Lett.} \textbf{\bibinfo{volume}{97}},
  \bibinfo{pages}{261106} (\bibinfo{year}{2010}).

\bibitem[{\citenamefont{Wacker et~al.}(2013)\citenamefont{Wacker, Lindskog, and
  Winge}}]{wacker2013nonequilibrium}
\bibinfo{author}{\bibfnamefont{A.}~\bibnamefont{Wacker}},
  \bibinfo{author}{\bibfnamefont{M.}~\bibnamefont{Lindskog}}, \bibnamefont{and}
  \bibinfo{author}{\bibfnamefont{D.}~\bibnamefont{Winge}},
  \bibinfo{journal}{IEEE J. Quantum Elect.} \textbf{\bibinfo{volume}{19}},
  \bibinfo{pages}{1200611} (\bibinfo{year}{2013}).

\bibitem[{\citenamefont{Fathololoumi et~al.}(2012)\citenamefont{Fathololoumi,
  Dupont, Chan, Wasilewski, Laframboise, Ban, M{\'a}ty{\'a}s, Jirauschek, Hu,
  and Liu}}]{fathololoumi2012terahertz}
\bibinfo{author}{\bibfnamefont{S.}~\bibnamefont{Fathololoumi}},
  \bibinfo{author}{\bibfnamefont{E.}~\bibnamefont{Dupont}},
  \bibinfo{author}{\bibfnamefont{C.}~\bibnamefont{Chan}},
  \bibinfo{author}{\bibfnamefont{Z.}~\bibnamefont{Wasilewski}},
  \bibinfo{author}{\bibfnamefont{S.}~\bibnamefont{Laframboise}},
  \bibinfo{author}{\bibfnamefont{D.}~\bibnamefont{Ban}},
  \bibinfo{author}{\bibfnamefont{A.}~\bibnamefont{M{\'a}ty{\'a}s}},
  \bibinfo{author}{\bibfnamefont{C.}~\bibnamefont{Jirauschek}},
  \bibinfo{author}{\bibfnamefont{Q.}~\bibnamefont{Hu}}, \bibnamefont{and}
  \bibinfo{author}{\bibfnamefont{H.}~\bibnamefont{Liu}}, \bibinfo{journal}{Opt.
  Express} \textbf{\bibinfo{volume}{20}}, \bibinfo{pages}{3866}
  (\bibinfo{year}{2012}).

\bibitem[{\citenamefont{Grange}(2014{\natexlab{a}})}]{grange2014SLs}
\bibinfo{author}{\bibfnamefont{T.}~\bibnamefont{Grange}},
  \bibinfo{journal}{Phys. Rev. B} \textbf{\bibinfo{volume}{89}},
  \bibinfo{pages}{165310} (\bibinfo{year}{2014}{\natexlab{a}}).

\bibitem[{\citenamefont{Grange}(2014{\natexlab{b}})}]{grange2014nanowire}
\bibinfo{author}{\bibfnamefont{T.}~\bibnamefont{Grange}},
  \bibinfo{journal}{Appl. Phys. Lett.} \textbf{\bibinfo{volume}{105}},
  \bibinfo{pages}{141105} (\bibinfo{year}{2014}{\natexlab{b}}).

\bibitem[{\citenamefont{Burghoff et~al.}(2012)\citenamefont{Burghoff, Chan, Hu,
  and Reno}}]{burghoff:261111}
\bibinfo{author}{\bibfnamefont{D.}~\bibnamefont{Burghoff}},
  \bibinfo{author}{\bibfnamefont{C.~W.~I.} \bibnamefont{Chan}},
  \bibinfo{author}{\bibfnamefont{Q.}~\bibnamefont{Hu}}, \bibnamefont{and}
  \bibinfo{author}{\bibfnamefont{J.~L.} \bibnamefont{Reno}},
  \bibinfo{journal}{Appl. Phys. Lett.} \textbf{\bibinfo{volume}{100}},
  \bibinfo{eid}{261111} (\bibinfo{year}{2012}).

\end{thebibliography}

\end{document}